\documentclass[prl,twocolumn,preprintnumbers,nofootinbib]{revtex4}
\usepackage[a4paper, hdivide={1.91cm,,1.165cm}, vdivide={1.83cm,,3.2cm}]{geometry}

\usepackage{amsmath,amssymb}
\usepackage{graphicx,multirow}
\usepackage{color}
\usepackage{units}
\usepackage[hyperfootnotes=false]{hyperref}

\def \L {\mathcal{L}} 

\def \vec#1{{\boldsymbol{#1}}}

\newcommand{\BR}{\ensuremath{\text{BR}}}
\newcommand{\del}{\partial}

\begin{document}

\title{Lepton Flavor Violation with Displaced Vertices}

\preprint{ULB-TH/17-18}

\author{Julian \surname{Heeck}}
\email{julian.heeck@ulb.ac.be}
\affiliation{Service de Physique Th\'eorique, Universit\'e Libre de Bruxelles, Boulevard du Triomphe, CP225, 1050 Brussels, Belgium}

\author{Werner \surname{Rodejohann}}
\email[Electronic address: ]{werner.rodejohann@mpi-hd.mpg.de}
\affiliation{Max-Planck-Institut f\"ur Kernphysik, Saupfercheckweg 1, 69117 Heidelberg, Germany}

\hypersetup{
    pdftitle={Lepton Flavor Violation with Displaced Vertices},
    pdfauthor={Julian Heeck, Werner Rodejohann}
}


\begin{abstract}
\noindent 
If light new physics with lepton-flavor-violating couplings exists, the prime discovery channel might not be $\ell\to\ell'\gamma$ but rather $\ell\to\ell' X$, where the new boson $X$ could be an axion, majoron, familon or $Z'$ gauge boson. The most conservative bound then comes from $\ell\to\ell'+\mathrm{inv}$, but if the on-shell $X$ can decay back into leptons or photons, displaced-vertex searches could give much better limits. We show that only a narrow region in parameter space allows for displaced vertices in muon decays, $\mu\to e X, X\to \gamma\gamma, ee$, whereas tauon decays can have much more interesting signatures.
\end{abstract}

\maketitle


\section{Introduction}

The Standard Model (SM) brings with it the accidental conservation of baryon number $B$ and individual lepton numbers $L_{e,\mu,\tau}$. The linear combination $B+\sum_\alpha L_\alpha$ is broken at a non-perturbative level~\cite{'tHooft:1976up} and the differences $L_\alpha -L_\beta$ are clearly violated by neutrino oscillations~\cite{Patrignani:2016xqp}. Despite of that, we have yet to observe a lepton-flavor-violating (LFV) process involving \emph{charged} leptons, which is, without additional assumptions, decoupled from neutrino oscillations and thus a perfect signature of new physics~\cite{Heeck:2016xwg}.

Assuming new particles with LFV couplings much heavier than the energies in question allows one to use an effective-field-theory approach with higher-dimensional operators, which typically make $\mu\to e \gamma$, $\mu\to 3e$ or $\mu$--$e$ conversion (Tab.~\ref{tab:lfv}) the best processes to detect LFV in the $\mu$--$e$ sector (see e.g.~Ref.~\cite{Lindner:2016bgg,Calibbi:2017uvl}). This conclusion no longer holds if new particles with LFV couplings exist that are \emph{lighter} than the muon. Examples for these are plentiful, be it light gauge bosons $Z'$~\cite{Foot:1994vd,Dobrescu:2004wz,McDonald:2006jr,Heeck:2014qea,Heeck:2016xkh,Altmannshofer:2016brv} or light (pseudo-)scalars, e.g.~familons, majorons or axion(-like) particles~\cite{Wilczek:1982rv,Reiss:1982sq,Kim:1986ax,Grinstein:1985rt,Pilaftsis:1993af,Feng:1997tn,Hirsch:2009ee,Jaeckel:2013uva,Celis:2014iua,Celis:2014jua,Ema:2016ops,Calibbi:2016hwq,Garcia-Cely:2017oco}. The usually considered processes listed above are then typically heavily suppressed, making the two-body decay $\mu\to e X$ the prime search channel. The signature here depends on the decay channels of $X$:
\begin{enumerate}
	\item If $X$ decays invisibly, for example into neutrinos or dark matter, only the mono-energetic electron can be searched for on top of the continuous Michel spectrum, with limits on $\BR (\mu\to e X)$ of order $10^{-5}$~\cite{Bayes:2014lxz} and $10^{-6}$~\cite{Jodidio:1986mz}, depending on $m_X$ and the chirality. The emission of an additional photon can help to further reduce background, leading to $\BR (\mu\to e \gamma X) < 10^{-9}$~\cite{Goldman:1987hy, Bolton:1988af}.
	\item If $X$ decays into visible particles, e.g.~$X\to e^+ e^-$ or $X\to \gamma\gamma$, much better limits could be obtained as long as the decay happens \emph{inside} of the detector. This typically involves a reconstruction of the displaced vertex (DV) of $X\to\mathrm{vis}$ and thus different cuts and triggers than usual. We stress that the signatures are background-free both due to their LFV nature and the DV. 
\end{enumerate}
Similar considerations hold for LFV $\tau$ decays, which allow for many more visible DV channels, including $X\to \text{hadrons}$. \emph{Invisible} $\tau\to \ell X$ decays have been studied at ARGUS~\cite{Albrecht:1995ht} (see also older limits in Refs.~\cite{Baltrusaitis:1985fh,Albrecht:1990zj}), and are under investigation at Belle~\cite{Yoshinobu:2017jti}.

\begin{ruledtabular}
\begin{table}[b]
	\centering
		\begin{tabular}{lll}
 Process & Current  & Future\\		
		\hline
 $\mu\to e$ conv. & $\mathcal{O}( 10^{-12})$~\cite{Bertl:2006up} & $10^{-17}$~\cite{Kuno:2013mha,Bartoszek:2014mya}\\
 $\mu\to e\gamma$ & $4.2\times 10^{-13}$~\cite{TheMEG:2016wtm} & $4\times 10^{-14}$\cite{Baldini:2013ke}\\
 $\mu\to e\bar e e$ & $1.0\times 10^{-12}$~\cite{Bellgardt:1987du} & $10^{-16}$~\cite{Blondel:2013ia}\\
 $\mu\to e \gamma \gamma$ & $7.2\times 10^{-11}$~\cite{Grosnick:1986pr} & -- \\
 $\mu\to e \gamma X, X\to \mathrm{inv}$ & $\mathcal{O}(10^{-9})$~\cite{Goldman:1987hy, Bolton:1988af} & -- \\
 $\mu\to e X, X\to \mathrm{inv}$ & $\mathcal{O}(10^{-5})$~\cite{Bayes:2014lxz} & $10^{-8}$~\cite{mu3e_familon}\\
 $\mu\to e X, X\to ee$ & $\mathcal{O}(10^{-11})$~\cite{Eichler:1986nj} & $< 10^{-14}$\\
 $\mu\to e X, X\to \gamma\gamma$ & $\mathcal{O}(10^{-10})$~\cite{Bolton:1988af,Natori:2012gga} & $10^{-11}$~\cite{Natori:2012gga}
		\end{tabular}
		\caption{LFV processes in the muon sector with current ($90\%$~C.L.) and future limits on branching ratios. Limits involving a new light boson $X$ depend on its mass, lifetime, and branching ratios, see references and text for details.
		\label{tab:lfv}}
\end{table}
\end{ruledtabular}

\begin{figure}[t]
\includegraphics[width=0.49\textwidth]{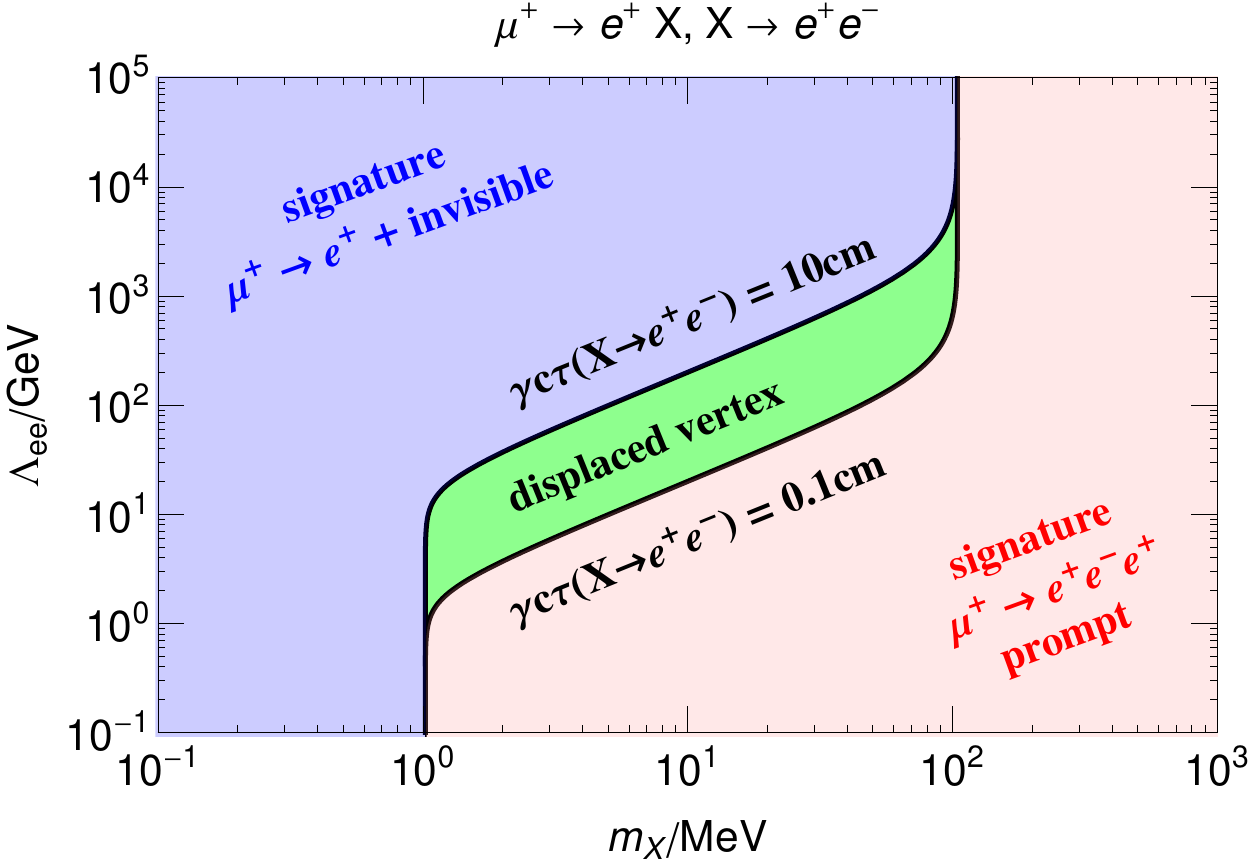}
\caption{
Signatures of $\mu\to e X$, $X\to ee$ depending on $X$ mass and $Xee$ coupling strength. For $m_X < 2 m_e$ or if $X$ decays outside of the detector, the signature is mainly $\mu\to e +\text{inv}$ (blue region). For $m_X>m_\mu-m_e$ or if the $X$ decay length $c\tau$ is smaller than the vertex resolution, the signature is just prompt $\mu\to 3e$ (red region). In between, a region with detectable displaced $X\to ee$ vertex exists (green), which of course depends on the detector geometry and acceptance.
}
\label{fig:phase_diagram_mu_eee}
\end{figure}

LFV decays \emph{with DV} are only possible in certain kinematical regions, e.g.~$2 m_e < m_X < m_\mu-m_e$ for $\mu\to eX, X\to ee$, and furthermore require the $X$ decay length to be larger than the experimental vertex resolution and smaller than the detector. This leaves a sliver of testable parameter space where limits can be put on $\BR(\mu\to e X, X\to ee)$, illustrated in Fig.~\ref{fig:phase_diagram_mu_eee} (see later for details). Since sub-GeV particles $X$ with couplings to leptons or photons are strongly constrained by other experimental searches, it is not obvious that there is viable parameter space for LFV DV. As we will see below, there is only a small feasible region for muon decays, whereas tauon decays are much less constrained and can have a plethora of interesting signatures.

The focus of this letter will be these LFV decays with DV. Existing work is scarce; we are not aware of any analyzes for $\tau$, but there is an old limit from SINDRUM on $\BR(\mu\to e X, X\to ee)$ of order $10^{-11}$~\cite{Eichler:1986nj} and a 
thesis within the MEG collaboration on $\BR(\mu\to e X, X\to \gamma\gamma)$ with a limit of order $10^{-11}$~\cite{Natori:2012gga}. We expect Mu3e~\cite{Blondel:2013ia} to  vastly improve at least the SINDRUM limit, and encourage searches for these kind of LFV $\tau$ decays at $B$ factories.

\section{Framework}

We focus our analysis on pseudo-Goldstone bosons $X$,\footnote{Vector bosons $Z'$ will behave similar to pseudoscalars since for light $Z'$ only the longitudinal component is produced in $\ell_\alpha \to \ell_\beta Z'$~\cite{Heeck:2016xkh}. The main difference is then in the $Z'$ decay, which in particular does not allow for $\gamma\gamma$.
Light CP-\emph{even} scalars look qualitatively similar and will typically mix with the Higgs boson, leading to additional couplings~\cite{Flacke:2016szy,Dev:2016vle,Dev:2017dui}.}  whose couplings to SM leptons $\ell_{e,\mu,\tau}$ can be conveniently parametrized as~\cite{Feng:1997tn}
\begin{align} \label{eq:L}
\L_X &= \frac{\del_\mu X}{\Lambda} \overline{\ell}_\alpha \gamma^\mu (g^V_{\alpha\beta} + g^A_{\alpha\beta}\gamma^5) \ell_\beta \\
&= -i\frac{X}{\Lambda} \overline{\ell}_\alpha (g^V_{\alpha\beta}(m_\alpha-m_\beta) + g^A_{\alpha\beta}(m_\alpha+m_\beta)\gamma^5) \ell_\beta\,,\nonumber
\end{align} 
with some effective scale $\Lambda$ and hermitian (anti-hermitian) coupling matrix $g^A$ ($g^V$), to be assumed real in the following. In the second line we have integrated by parts and used the equations of motion, which is justified for on-shell particles. 
In the case of leptonic familons~\cite{Wilczek:1982rv,Reiss:1982sq,Kim:1986ax}, $\Lambda$ corresponds to the scale of the broken global flavor symmetry and the matrix structure of $g^{A,V}$ is determined by the symmetry generators~\cite{Feng:1997tn}. However, these couplings arise even in simple unflavored singlet-majoron models at one-loop level~\cite{Pilaftsis:1993af} and depend on seesaw parameters~\cite{Garcia-Cely:2017oco}; in fact, measuring these majoron couplings could make it possible to reconstruct the seesaw parameters without having to detect the heavy neutrinos. Diagonal as well as off-diagonal couplings to charged leptons are thus a generic part of many models and in particular relevant for neutrino-mass models with global symmetries.
	
We assume the mass of $X$, $m_X$, to be an independent parameter.
It proves convenient to define the scales 
\begin{align}
\Lambda_{\alpha\beta}\equiv \Lambda/\sqrt{(g^V_{\alpha\beta})^2+(g^A_{\alpha\beta})^2}\,.
\end{align}
The LFV two-body decays then take the form
\begin{align}
\Gamma &(\ell_\alpha \to \ell_\beta X) = \frac{m_\alpha^3}{16\pi \Lambda^2} \sqrt{\left(1-r_X^2\right)^2+r_\beta^4-2 r_\beta^2 \left(1+r_X^2\right)}\nonumber\\ 
&\quad\times\left[\left((g^A_{\alpha\beta})^2+(g^V_{\alpha\beta})^2\right) \left(1-r_\beta^2\right)^2\right.\\
&\quad\quad\left.-\left((g^V_{\alpha\beta})^2 (1-r_\beta)^2+(g^A_{\alpha\beta})^2 (1+r_\beta)^2\right) r_X^2\right] ,\nonumber
\label{eq:familon_emission}
\end{align}
with $r_{\beta,X} \equiv m_{\beta,X}/m_\alpha$.
For $m_{\beta,X}\ll m_\alpha$, this is simply $m_\alpha^3/(16\pi \Lambda_{\alpha\beta}^2)$.
The boson decay is given by
\begin{align}
&\Gamma (X\to \ell_\alpha \bar \ell_\alpha) = \frac{ m_X}{2\pi} (g^A_{\alpha\alpha})^2\frac{m_\alpha^2}{\Lambda^2} \sqrt{1-\frac{4m_\alpha^2}{m_X^2}}\,,\\
&\Gamma (X\to \ell_\alpha \bar \ell_\beta+\bar \ell_\alpha \ell_\beta) \simeq \frac{m_X}{4\pi}\frac{m_\alpha^2}{\Lambda_{\alpha\beta}^2} \left(1-\frac{m_\alpha^2}{m_X^2}\right)^2 ,
\end{align}
the last equation being valid for $m_\beta\ll m_\alpha$. 

The decay $X\to\gamma\gamma$ induced by a fermion loop is typically suppressed, but of course becomes the dominant decay channel for $m_X<2 m_e$~\cite{CorderoCid:2005gr}. We will simply assume an effective photon coupling~\cite{Patrignani:2016xqp},
\begin{align}
\L\supset -\frac{g_{\gamma\gamma}}{4} X F_{\mu\nu}\tilde F^{\mu\nu} && \Rightarrow &&\Gamma (X\to\gamma\gamma) = \frac{g_{\gamma\gamma}^2 m_X^3}{64\pi}\,,
\end{align}
with field-strength tensor $F_{\mu\nu}$ and its dual $\tilde F^{\mu\nu}=\tfrac12 F_{\alpha\beta} \epsilon^{\mu\nu\alpha\beta}$.
The coupling $g_{\gamma\gamma}$ with mass dimension $-1$ could be generated by a triangle anomaly analogous to axions or via mixing with the longitudinal $Z$ component as in majoron models~\cite{Garcia-Cely:2017oco,Heeck:2017xbu}.
In addition to the decay into leptons and photons one could easily imagine invisible decays (into neutrinos or additional new light particles) or decays into hadrons (for sufficiently heavy $X$). To simplify the analysis we will neglect these channels. 

The relevant quantity for DV is the decay length in the laboratory frame, in which $X$ is typically boosted. For LFV with muons (e.g.~MEG or Mu3e), the muon is stopped before it decays into $e X$, so $X$ has the following momentum in the lab frame
\begin{align}
\hspace{-1ex} |\vec{p}_X| = \frac{\sqrt{(m_\mu^2 - (m_e+m_X)^2)(m_\mu^2 - (m_e-m_X)^2)}}{2 m_\mu} \,,
\end{align}
leading to the boosted decay length~\cite{Patrignani:2016xqp}
\begin{align}
\gamma c \tau = \frac{ c |\vec{p}_X|}{m_X \Gamma_X}\,.
\label{eq:boosted_decay_length}
\end{align}
Now $P(x) = \exp (- x/\gamma c \tau)$ is the probability for $X$ to travel a distance $x$ without decaying. Note that the inclusion of additional $X$ decay channels can only \emph{shorten} the decay length, rendering the decay more prompt and reducing the rate by $1-\BR(X\to \text{inv})$.

For tau decays (e.g.~in Belle or LHCb) the situation is more complicated because the particles do not decay at rest in the lab frame. We will leave a dedicated analysis to our experimental colleagues but nevertheless approximate the tau at rest in the following. The additional boost can increase or decrease the physical decay length, depending on the direction of $X$ emission in the tau frame. Since we will see that the parameter space for tau decays is wide open, our conclusions should be qualitatively correct.

\begin{figure*}[t]
\includegraphics[width=0.49\textwidth]{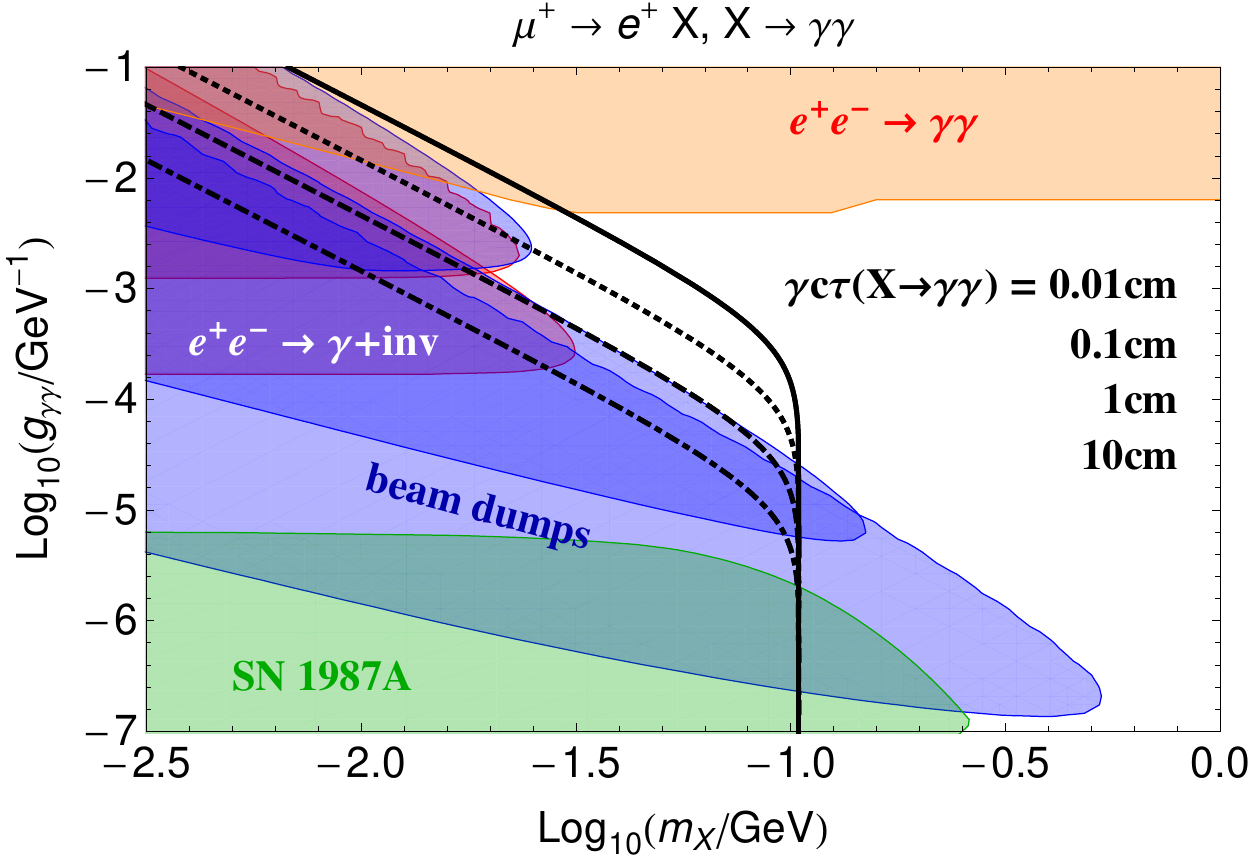}
\includegraphics[width=0.49\textwidth]{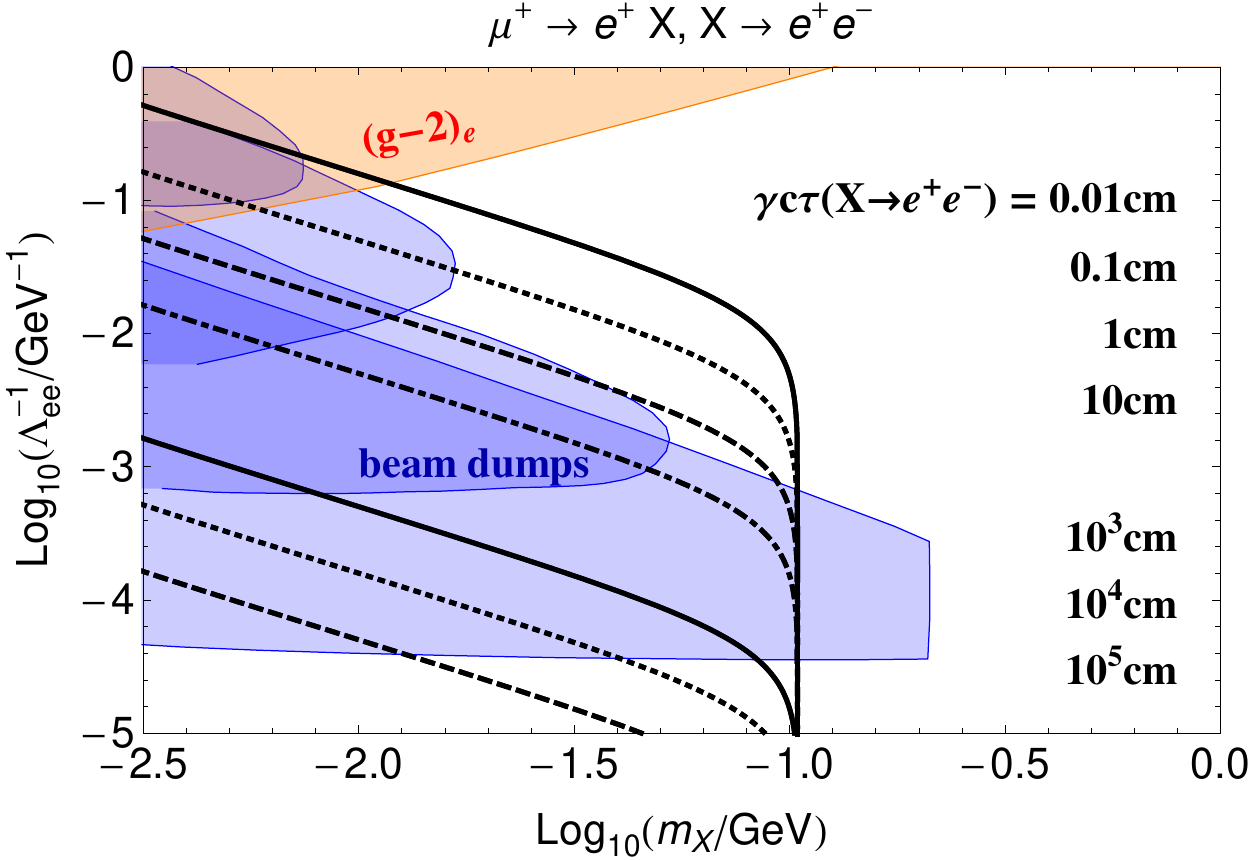}\\[2ex]
\includegraphics[width=0.49\textwidth]{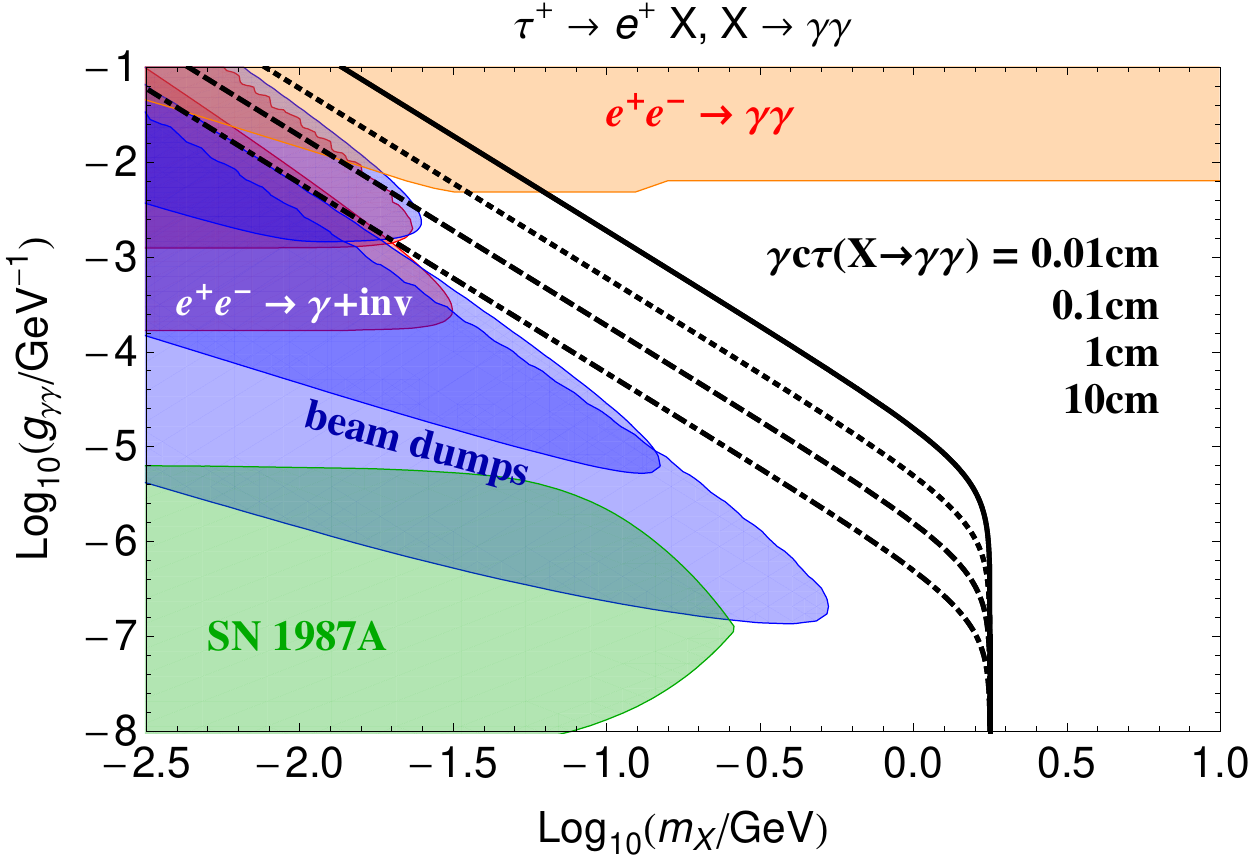}
\includegraphics[width=0.49\textwidth]{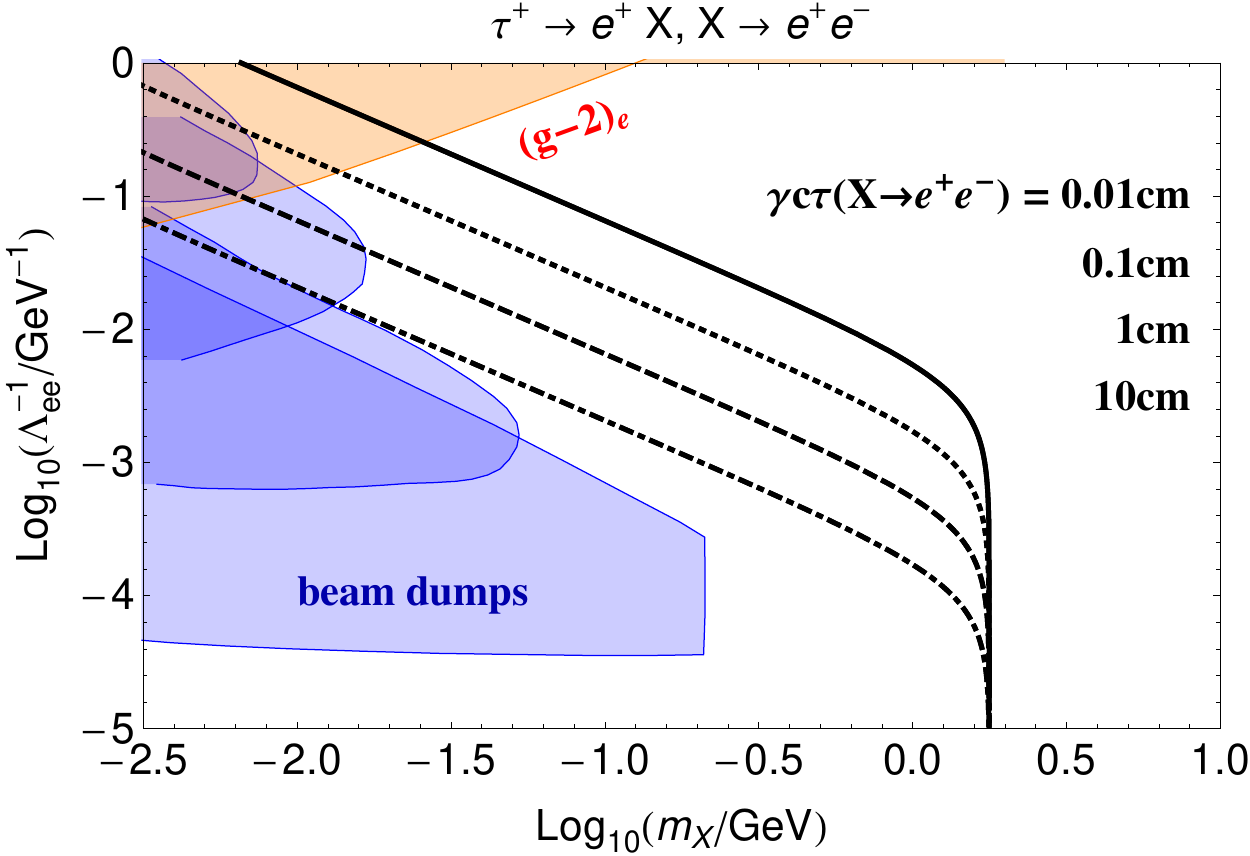}
\caption{
Left: excluded regions for a scalar $X$ with mass $m_X$ and coupling to photons $g_{\gamma\gamma}$~\cite{Jaeckel:2015jla,Dobrich:2015jyk,Dolan:2017osp}. In black we show contours of the boosted decay length $\gamma c\tau$ of $X\to\gamma\gamma$, assuming $X$ to be produced from an at-rest muon decay $\mu \to e X$ (upper panel) or tauon decay $\tau \to e X$ (lower panel). Here, the solid black line corresponds to $\gamma c\tau = \unit[0.01]{cm}$, the dotted one to $\unit[0.1]{cm}$, the dashed one to $\unit[1]{cm}$ and the dot-dashed line to $\unit[10]{cm}$.
Right: same as left, but for a scalar $X$ with coupling to electrons, so the contours are for the boosted decay length $\gamma c\tau$ of $X\to ee$.
}
\label{fig:LFV_DV}
\end{figure*}

\section{Muon decays}

We start our analysis with LFV muon decays, which kinematically allow for $\mu\to eX$ with $X\to \gamma\gamma$ or $X\to ee$.

\subsection{\texorpdfstring{Decay $\mu\to e X, X\to \gamma\gamma$}{mu to eX, X to gamma gamma}}

Assuming only $g_{e\mu}$ and $g_{\gamma\gamma}$ to be non-zero, we have the branching ratio in the narrow-width approximation~\cite{CorderoCid:2005gr}
\begin{align}
\BR(\mu\to e X, X\to \gamma\gamma) &\simeq \BR(\mu\to e X)\BR (X\to\gamma\gamma) \nonumber\\
&\simeq \BR(\mu\to e X)\,,
\label{eq:BRNWA}
\end{align}
and the boosted decay length from Eq.~\eqref{eq:boosted_decay_length},
\begin{align}
\hspace{-1ex} \gamma c \tau \simeq \frac{32\pi m_\mu}{g_{\gamma\gamma}^2 m_X^4}
\simeq \unit[21]{cm} \left(\frac{\unit{TeV^{-1}}}{g_{\gamma\gamma}}\right)^2\left(\frac{\unit[10]{MeV}}{m_X}\right)^4.
\label{eq:decay_length_gammagamma}
\end{align}
The experiment of choice for this decay chain is MEG~\cite{Natori:2012gga} due to the better photon detection compared to Mu3e.\footnote{This could change with the extension of Mu3e by a photon conversion layer~\cite{Cheng:2013paa,mu3e_familon}.} While MEG's detector geometry should allow for reconstructed vertices up to the meter scale, we can see from Fig.~\ref{fig:LFV_DV} (upper left) that such large decay lengths are incompatible with beam dump data. Limits on $g_{\gamma\gamma}$, re-derived and updated recently in Refs.~\cite{Jaeckel:2015jla,Dobrich:2015jyk,Dolan:2017osp}, are in fact so strong that they exclude $X$ masses below $\unit[20]{MeV}$ and decay lengths longer than cm. Future experiments such as NA62, Belle-II, and SHiP~\cite{Dobrich:2015jyk,Dolan:2017osp} can push this limit to $\unit[0.1]{cm}$ (see also Ref.~\cite{Bauer:2017ris} for LHC prospects). 

While vertex resolutions of order cm might be possible in MEG(-II), many of the decays will appear prompt, but still have a different energy distribution from the general three-body decay $\mu\to e\gamma\gamma$. For not-too-heavy $X$, the positron energy will actually be similar to that from $\mu\to e\gamma$ for which MEG is optimized, which should improve the efficiency of this search.
Assuming an improvement of the 30-year-old Crystal-Box limit~\cite{Bolton:1988af} by an order of magnitude with MEG(II), i.e.~a reach down to $\BR(\mu\to e X, X\to \gamma\gamma)  \simeq 10^{-11}$ for sufficiently prompt $X$ decays~\cite{Natori:2012gga}, this corresponds to LFV scales $\unit[10^{12}]{GeV}\lesssim\Lambda_{\mu e}$. For comparison, $\BR(\mu\to e X)$ with invisible $X$ decay currently gives a lower limit $\unit[10^{9}]{GeV}\lesssim\Lambda_{\mu e}$; if this were to be improved to $\BR(\mu\to e +\text{inv}) \lesssim 10^{-8}$ with Mu3e~\cite{mu3e_familon} one could push this to $\unit[10^{11}]{GeV}\lesssim\Lambda_{\mu e}$.
This illustrates nicely how much limits on $\BR(\mu\to e X)$ can be improved if $X$ decays back into observable particles within the detector.

Optimistically, the observation of LFV DV allows us to determine three quantities: the invariant $\gamma\gamma$ mass gives $m_X$, the total rate $\mu\to e X, X\to\gamma\gamma$ gives $\Lambda_{\mu e}$ via Eq.~\eqref{eq:BRNWA}, and the decay length gives $g_{\gamma\gamma}$ via Eq.~\eqref{eq:decay_length_gammagamma}, i.e.~the region in Fig.~\ref{fig:LFV_DV}.
This is the reason why LFV DV is such an interesting signature to pursue.

\subsection{\texorpdfstring{Decay $\mu\to e X, X\to ee$}{mu to eX, X to ee}}

Setting all $X$ couplings but $g_{e\mu}$ and $g_{ee}$ to zero allows us to determine the $X$ decay length of $\mu\to e X, X\to ee$ from Eq.~\eqref{eq:boosted_decay_length},
\begin{align}
\hspace{-2ex} \gamma c \tau \simeq \frac{\pi m_\mu \Lambda_{ee}^2}{m_e^2 m_X^2}
\simeq \unit[2.5]{cm} \left(\frac{\Lambda_{ee}}{\unit[100]{GeV}}\right)^2\left(\frac{\unit[10]{MeV}}{m_X}\right)^2 \hspace{-0.5ex} ,
\end{align}
and compare to existing limits on $g_{ee}$. 
At one loop, $X$ contributes negatively to leptonic magnetic moments~\cite{Essig:2010gu}, so we can obtain a bound from $(g-2)_e$~\cite{Davoudiasl:2014kua}. We will not bother deriving collider constraints on $g_{ee}$ (e.g.~$e^+ e^-\to \gamma X, X\to e^+ e^-$~\cite{Lees:2014xha}) because they are not relevant for our region of interest.
The most important constraints come once more from beam dumps~\cite{Andreas:2010ms,Essig:2010gu}, which again prohibit decay lengths longer than cm, see Fig.~\ref{fig:LFV_DV} (upper right).
Note that Mu3e should be able to set a limit on $g_{ee}$ via $\mu^+\to e^+\bar{\nu}_\mu\nu_e X, X\to e^+e^-$ without LFV, analogous to the dark photon case discussed in Ref.~\cite{Echenard:2014lma}.

Similar to the diphoton decay, $\mu\to e X, X\to ee$ with a DV around cm could potentially be distinguished from prompt decays at future experiments such as Mu3e, but this requires a dedicated analysis.
The $X$ mass is necessarily large in this region; for instance, 
from Fig.~\ref{fig:LFV_DV} one can read off that decay lengths below 1 mm correspond to allowed masses $m_X \gtrsim \unit[15]{MeV}$.
Since the decay length is rather short, many of the decays will pass the cuts for prompt $\mu\to 3e$. The light-physics origin can nevertheless leave a trace in the Breit--Wigner $X$ peak of the invariant $e^+e^-$ mass. This is of course a very optimistic scenario in which we observe so many LFV events that we can determine the differential distributions.

Mu3e aims to improve the BR limit on prompt $\mu\to 3e$ decays down to $10^{-16}$~\cite{Blondel:2013ia}; prompt-enough $\mu\to e X, X\to ee$ decays should then naively be probed well below $10^{-14}$, which corresponds to limits on $\Lambda_{\mu e}$ up to $\unit[8\times 10^{13}]{GeV}$ if $\BR(X\to ee)\simeq 1$. This is the highest testable LFV scale in our analysis. 

With non-zero couplings $g_{e\mu}$ and $g_{ee}$ the boson $X$ unavoidably contributes to $\mu\to e \gamma$ at loop level. Defining the function
\begin{align}
f(x)\equiv 1- 2 x + 2 x (x-1) \log\left(\frac{x}{x-1}\right) ,
\end{align}
which is $1 + \mathcal{O}(x)$ for small $x$, the $\mu\to e\gamma $ branching ratio takes the simple form~\cite{Lavoura:2003xp}
\begin{align}
\BR (\mu\to e\gamma)&\simeq \frac{3\alpha m_e^2}{8\pi m_\mu^2 G_F^2 \Lambda_{ee}^2 \Lambda_{\mu e}^2} \left|f\left(\frac{m_X^2}{m_\mu^2}\right)\right|^2\\
&\sim 10^{-18}\left(\frac{\unit[10]{GeV}}{\Lambda_{ee}}\right)^2\left(\frac{\unit[10^9]{GeV}}{\Lambda_{\mu e}}\right)^2,
\end{align}
to lowest order in the electron mass. This is unobservably small for the values probed by $\mu\to e X$ and $\mu\to e X, X\to ee$, illustrating the importance of light-boson searches.

For $\mu\to e X, X\to ee$ there is potentially a second region of interest, with $g_{ee}$ couplings \emph{below} the beam-dump limits. This region was excluded by the constraints from the supernova SN1987A for the di-photon channel (Fig.~\ref{fig:LFV_DV} (upper left)), but the situation is different here. Supernova limits on $g_{ee}$ have of course been derived early on~\cite{Mayle:1987as}, but usually in the context of axions where the coupling to quarks is dominant. Thus, while supernova constraints have been significantly improved and refined for most other light-new-physics models and couplings~\cite{Hardy:2016kme}, there has been surprisingly little progress for $g_{ee}$. While we naively expect the supernova limit to overlap with the beam-dump limits as in most cases, the recent evaluation of the scalar coupling $X\overline{e}e$ in Ref.~\cite{Knapen:2017xzo} shows that a gap between them is also possible.
An updated constraint on our $g_{ee}$ following for example Ref.~\cite{Hardy:2016kme} goes beyond the scope of our letter but is certainly a worthwhile endeavor. Let us for now assume that there is viable parameter space around $\Lambda_{ee}\sim\unit[30]{TeV}$, i.e.~just below the beam-dump limits. This corresponds to decay lengths above $\unit[10^3]{cm}$, which is of course far outside of the detector. Nevertheless, the probability for $X$ to decay within the detector is not necessarily very small, roughly $1-P(x)\simeq x/\gamma c\tau$. The effective branching ratio for $X$ decay inside the detector is then 
\begin{align}
\BR(\mu\to e X) &\BR (X\to ee) (1-P(l_\text{dec})) \nonumber\\
&\simeq \BR(\mu\to e X) \frac{l_\text{dec}}{\gamma c \tau} \,.
\end{align}
Now, $\BR(\mu\to e X)$ is expected to be pushed down to $10^{-8}$ in Mu3e~\cite{mu3e_familon}, while $l_\text{dec}/\gamma c\tau$ can be as big as $10^{-3}$. Thus, effective branching ratios $\BR(\mu\to e X, X\to ee)$ \emph{with a DV} in the detector can be as big as $10^{-11}$.
Compared to the case of rather short-lived $X$ discussed before, very few of the $X\to ee$ decays will appear prompt here, with most decays at the edge of the detector. This will reduce the efficiency of the search, but should still allow to improve the limit below $10^{-11}$ with a dedicated analysis.
If $\mu\to eX$ is observed, a search for $\mu\to e X, X\to ee$ becomes of course obligatory.
Pending updated supernova constraints, there is room for LFV DV $\mu\to e X, X\to ee$ anywhere in the Mu3e detector.

\section{Tauon decays}

The same analysis as before can be made for tauons, with current LFV limits coming mostly from BaBar and Belle, about to be improved with LHCb and Belle-II~\cite{Heeck:2016xwg}.

\subsection{\texorpdfstring{Decay $\tau\to \ell X, X\to \gamma\gamma$}{tau to l X, X to gamma gamma}}

In complete analogy to the muon case, we can study $\tau\to \ell X, X\to \gamma\gamma$ by setting all other $X$ couplings to zero. Note that the cases $\ell = e$ and $\ell = \mu$ differ only near the phase-space closure $m_X\simeq m_\tau-m_\ell$.
As can be seen from Eq.~\eqref{eq:boosted_decay_length}, the $X$ decay length is boosted by a factor $m_\tau/m_\mu$ compared to the $\mu\to e X$ decay, easily allowing for decay lengths of order $\unit[10]{cm}$. In addition, the kinematically accessible masses $m_X\sim \unit{GeV}$ can evade beam-dump limits altogether and essentially allow for arbitrarily long or short $X$ decay lengths (Fig.~\ref{fig:LFV_DV} lower left).
Thus, even without proper knowledge of the tau momentum distribution in the lab frame or the vertex resolution of Belle or LHCb we can confirm the possibility of LFV DV tauon decays and encourage a dedicated search.

We are not aware of any experimental searches for $\tau\to\ell \gamma\gamma$, but limits of order $\BR \lesssim 10^{-7}$ exist on the LFV channels $\tau\to \ell M^0, M^0\to\gamma\gamma$, with $M^0\in \{\pi^0,\eta^0\}$~\cite{Miyazaki:2007jp}. We can expect limits of the same order for our $X$-mediated channels, which corresponds to scales $\Lambda_{\tau\ell}\sim \unit[5\times 10^8]{GeV}$. For comparison, old ARGUS limits on the invisible channel $\BR (\tau\to \ell X) \lesssim 5\times 10^{-3}$~\cite{Albrecht:1995ht} correspond to $\Lambda_{\tau\ell}\gtrsim \unit[2\times 10^6]{GeV}$. Also these limits will be improved with Belle~\cite{Yoshinobu:2017jti}, but can probably never reach the same sensitivity as $\tau\to \ell X, X\to \gamma\gamma$.

\subsection{\texorpdfstring{Decay $\tau\to \ell X, X\to ee$}{tau to l X, X to ee}}

Just like for $\tau\to \ell X, X\to \gamma\gamma$, the large tauon mass boosts the $\tau\to \ell X, X\to ee$ decay length out of the highly-constrained beam-dump region, allowing once more essentially arbitrary DV.
Current limits on the prompt decays $\tau\to \ell ee$ are of order $10^{-8}$~\cite{Patrignani:2016xqp}, which corresponds to scales $\Lambda_{\tau\ell}\sim \unit[10^9]{GeV}$ if the $X$ decay is sufficiently prompt and $\BR(X\to ee)\simeq 1$. The sensitivity should not suffer much for the case of DV, as long as most $X$ decay within the detector.

\subsection{\texorpdfstring{Decay $\tau\to \ell X, X\to \mu\ell'$}{tau to l X, X to mu l'}}

The large tauon mass allows for a plethora of kinematically possible $X$ decays. Focusing on the muon decay $X\to\mu\mu$, the main constraint on that coupling comes from the muon's magnetic moment. At one loop, $X$ contributes with a negative sign to $(g-2)_\mu$~\cite{Essig:2010gu}, whereas the experimental value is infamously larger than the SM prediction by about $3\sigma$. Using very conservatively the $5\sigma$ constraint from $(g-2)_\mu$, we obtain a lower limit $\Lambda_{\mu\mu}\simeq \unit[500]{GeV}$ ($\unit[93]{GeV}$) for $m_X \ll m_\tau$ ($m_X=\unit[1.5]{GeV}$).
This poses no problem for LFV DV, seeing as a large decay length requires much larger $\Lambda_{\mu\mu}$:
\begin{align}
\gamma c\tau &\simeq \frac{\pi  m_\tau \Lambda_{\mu\mu}^2 (1-m_X^2/m_\tau^2)}{m_\mu^2 m_X^2 \sqrt{1-4m_\mu^2/m_X^2}}\\
&\simeq \unit[10]{cm}\left(\frac{\unit{GeV}}{m_X}\right)^2\left(\frac{\Lambda_{\mu\mu}}{\unit[10^6]{GeV}}\right)^2\frac{(1-m_X^2/m_\tau^2)}{\sqrt{1-4m_\mu^2/m_X^2}} .\nonumber
\end{align}
The parameter space for LFV with muon DV is thus wide open and ready to be explored. We expect the same sensitivity as for the $ee$ mode discussed above.

The last remaining LFV decay with displaced (neutral) vertex with leptons is $\tau\to e X, X\to \mu e$. Since $m_\mu + m_e < m_X$ by construction, there are no $\mu \to eX$ constraints to limit $\Lambda_{\mu e}$, so the main constraint comes from $(g-2)_\mu$ again, which is weak. It is hence completely possible to have $\tau^+\to e^+ X, X\to \mu^\mp e^\pm$ with a DV deep inside the detector, which has ample observables to identify it and should be a very background free decay.

\section{Conclusion}

The search strategies for rare lepton-flavor-violating decays have historically usually been motivated by heavy new physics, allowing for the use of effective field theory. While this indeed covers an immense region of model space, the absence of any signal so far implores us to challenge this approach. An obvious loophole comes in the form of light new particles with flavor violating couplings, which could be produced on-shell and travel a finite distance in the detector before decaying back into known particles. Here we have shown that such LFV displaced vertices are indeed possible for tauon decays $\tau \to \ell X$, $X\to \ell' \ell'',\gamma\gamma$, with essentially unconstrained decay length. For muon decays $\mu \to e X$, $X\to \gamma\gamma$, beam-dump experiments and supernova data already constrain the decay length to be below a cm, rendering these decays fairly prompt. The $\mu \to e X$, $X\to e e$ mode requires a dedicated re-analysis of supernova limits to evaluate the potential displaced-vertex lengths.
We urge our experimental colleagues to perform dedicated searches for rare LFV DV decay channels, e.g.~at MEG(-II), Mu3e, and Belle(-II).

\section*{Acknowledgements}
We thank Martin Bauer, Ann-Kathrin Perrevoort, and especially Andr\'{e} Sch\"oning for helpful discussions. JH is a postdoctoral researcher of the F.R.S.-FNRS. WR is supported by the DFG with grant RO 2516/6-1 in the Heisenberg program.

\bibliographystyle{utcaps_mod}
\bibliography{BIB}

\end{document}